\documentclass[]{interact}

\usepackage{epstopdf}

\usepackage[numbers,sort&compress]{natbib}
\bibpunct[, ]{[}{]}{,}{n}{,}{,}
\renewcommand\bibfont{\fontsize{10}{12}\selectfont}
\makeatletter
\def\NAT@def@citea{\def\@citea{\NAT@separator}}
\makeatother

\theoremstyle{plain}

\theoremstyle{definition}

\theoremstyle{remark}

\usepackage{graphicx} 
\usepackage{amsmath}
\usepackage{booktabs}
\usepackage{amsfonts}
\usepackage{subcaption}
\usepackage{hyperref}
\usepackage{tikz}
\usepackage[ruled,vlined]{algorithm2e}
\usepackage{relsize} 
\newcommand{\TCS}[2][]{\begin{tikzpicture}[baseline,#1]
	\foreach \X [evaluate=\X as \Y using {int(mod(\X,5))}]in {1,...,#2}
	{\ifnum\Y=0
		\draw (\X*0.5ex+0.3ex,0) -- ++(-2.8ex,2ex);
		\else
		\draw (\X*0.5ex+0.3ex,0) -- ++(-0.3ex,2ex);
		\fi}
	\end{tikzpicture}}
\tikzstyle{arrow} = [thick,->,>=stealth]

\begin{document}

\articletype{}

\title{A Bayesian Bi-Directional Splitting Framework for Variable Selection in Large Datasets}

\author{
\name{Aaron Coats\textsuperscript{a}*\thanks{*Email: a.coats.1@research.gla.ac.uk}, Vinny Davies\textsuperscript{a} and Mayetri Gupta\textsuperscript{a}}
\affil{\textsuperscript{a}School of Mathematics and Statistics, University of Glasgow.}
}

\received{}
\maketitle

\begin{abstract}
Modern tabular datasets are becoming increasingly large, both in the number of samples and covariates, posing significant challenges for Bayesian variable selection due to the resulting computational burden. While there is extensive literature on scaling Bayesian inference to large numbers of observations or high-dimensional covariate spaces, comparatively little work addresses scalable Bayesian variable selection when both dimensions are large simultaneously in a practical setting. This paper presents a novel Bayesian variable selection framework for efficiently analysing data with a large number of both rows and columns. The proposed framework operates via a divide-and-conquer approach, splitting data into batches along both directions and analysing each batch independently in parallel, after which the results are combined together in a two-phase consensus procedure. Experiments demonstrate the computational gains while retaining strong variable selection performance, successfully identifying relevant signals in noisy, high-dimensional settings despite the loss of information induced by data splitting. Practical guidelines are also provided, including recommendations for tuning parameter choices and effective data partitioning strategies, followed by a real application to an H3N2 influenza dataset.
\end{abstract}

\begin{keywords}
Bayesian computation; divide-and-conquer; variable selection; parallel computing; Monte Carlo
\end{keywords}

\section{Introduction}
Modern statistical datasets are increasingly characterised by both large numbers of observations and large numbers of covariates, creating substantial computational challenges for regression modelling and variable selection. Such high-dimensional structures arise in diverse settings including economics \citep{kalina2017high},  genome-wide association studies \citep{Uffelmann2021}, and  studies of pathogen evolution such as avian influenza dynamics \citep{h3n2}. In these settings, classical regression and Bayesian variable selection approaches become computationally prohibitive due to the cost of repeated operations on large data matrices \citep{davies2017sparse}, motivating the development of scalable inference methods tailored to modern high dimensional regimes. \\

Consider a regression setting with tabular data consisting of $n$ observations and $p$ candidate predictors. Such models provide a flexible framework for understanding the relationship between a response variable and a potentially large collection of explanatory covariates, and naturally support variable selection through identification of a subset of relevant predictors. However, as both $n$ and $p$ grow large, standard regression and variable selection procedures become computationally demanding due to repeated operations involving large design matrices, while storage constraints may prevent the full dataset from being handled on a single machine. Further challenges arise in high dimensional settings where $p>n$, which leads to non identifiable coefficient estimates without additional regularisation, and in situations where predictors exhibit strong collinearity, making reliable variable selection more difficult. \\

Attempts to perform accurate variable selection in high dimensional settings have been widely studied in the statistical literature. Penalised shrinkage methods such as ridge regression \citep{ridge}, LASSO \citep{lasso} and elastic net \citep{elastic_net} are widely used because they provide computationally efficient regularisation and can perform variable selection even when $p>n$. In particular, the elastic net is designed to improve stability in the presence of strongly correlated predictors. However, these approaches do not naturally incorporate uncertainty in the selection process, and typically rely on optimisation procedures that return a single regularised solution. In contrast, Bayesian variable selection (BVS) methods incorporate sparsity assumptions through prior distributions such as spike-and-slab formulations \citep{bvs_handbook}, allowing uncertainty in regression coefficients to inform the identification of relevant predictors. Despite these advantages, posterior inference in Bayesian variable selection models remains computationally demanding, particularly when relying on sampling-based methods such as Markov chain Monte Carlo (MCMC), motivating the development of scalable approaches tailored to modern datasets. \\

Various techniques have been approached to tackle the computational difficulties induced by large datasets. Data subsampling approaches \citep{Quiroz2014, Maclaurin2014}, for example, reduce computational cost by evaluating the posterior density using only a fraction of the observations at each step of an MCMC sampler. However, subsampling does not address the `large $p$' problem, and and typically requires the full dataset to remain accessible, which can introduce storage constraints. In addition, such methods are inherently serial in their implementation, limiting opportunities for parallelisation. \\

An alternative strategy is provided by Divide-and-Conquer (D\&C) approaches \citep{guo_2021, madasub}, which address computational limitations by distributing the analysis across multiple CPU cores. In these methods, the dataset is partitioned into subsets (or batches), each of which is analysed independently in parallel before the batch-level outputs are combined to form a consensus estimate. From a computational perspective, it is particularly advantageous to be able to adopt an embarrassingly parallel structure \citep{neiswanger2014}, in which communication between cores is avoided, as this minimises overhead while retaining scalability. The ability to operate on smaller subsets while exploiting parallel computation makes D\&C strategies especially attractive for Bayesian variable selection, although constructing a reliable consensus set of selected variables from the batch-level results remains a key challenge.\\

Current literature on merging strategies largely reflects the two main ways in which D\&C methods have been applied to tabular datasets: partitioning the rows and partitioning the columns. Row-splitting approaches have primarily been developed for scalable posterior inference in regression settings. For example, \citet{consensus} construct a consensus posterior by taking a weighted average of MCMC draws of regression coefficients obtained from each batch. Similarly, \citet{nemeth2018merging} model the relationship between sub posterior draws and their corresponding log sub posterior densities using Gaussian processes, which are then used to construct approximate draws from the full data posterior. Extending this framework further, \citet{Buchholz_2023} consider row-wise partitioning for inference on the model posterior itself rather than only the regression coefficients. While effective for large $n$, these approaches are not designed specifically for variable selection. \\

Approaches based on column-wise partitioning have instead been developed more directly in the context of variable selection. For example, \citet{liang_2015} perform BVS by applying a screening procedure independently to each batch and aggregating the resulting selections to obtain a consensus set of predictors. A similar screening process is used in \citet{basil} for variable selection in the context of the LASSO. However, these approaches address scalability primarily with respect to the number of covariates, and operate in only a single partitioning direction. Methods that combine row-wise and column-wise partitioning within a unified D\&C framework for variable selection remain comparatively underexplored. \\

This paper presents a novel D\&C framework for variable selection that partitions a dataset along both rows and columns to improve computational efficiency in settings where both the number of observations and covariates are large. The proposed framework begins by forming batches using one of two partitioning strategies. A screening procedure akin to \citet{liang_2015} is then applied within each batch to identify variables considered significant, producing a set of candidate predictors for each row split. Finally, this information is merged across rows partitions using consensus techniques adapted from the row-splitting literature, producing a final set of selected predictors. \\

The remainder of this paper is organised as follows: Sections \ref{sec:background_tools} and \ref{sec:background_methods} provides a review of relevant background material, before the proposed framework is detailed in Section \ref{sec:our_method}. Section \ref{sec:simulations} presents computer experiments demonstrating the advantages of the proposed framework and provides recommendations for its practical implementation. A real data case study identifying antigenic sites in the H3N2 influenza virus is then presented in Section \ref{sec:h3n2}, followed by concluding remarks in Section \ref{sec:conclusions}.

\section{Background}
\label{sec:background_tools}
First, a discussion on some of the statistical tools relevant to the BVS framework proposed in Section~\ref{sec:our_method} is presented. Consider a univariate vector of responses, $\textbf{y} \in \mathbb{R}^n$, and a design matrix of covariate values, $\textbf{X} = [\textbf{x}_1, \dots, \textbf{x}_n]^\top \in \mathbb{R}^{n \times p}$. The response is modelled as $\textbf{y} \sim N(\textbf{X}\boldsymbol{\beta}, \sigma^2 \mathbb{I}_n)$, where $\boldsymbol{\beta}$ is a vector of unknown regression coefficients $\beta_j$ of length $p$, and $\sigma^2 \in \mathbb{R}^+$ is the residual variance. The goal of variable selection is to identify the subset of $\{1,\dots , p\}$ associated with the variables most strongly linked to the response.

\subsection{Sparsity-inducing Priors}
\label{sec:spike_slab}
In a high-dimensional setting, model interpretability is enhanced through the assumption of sparsity, which aims to drop as many noisy covariates as possible, preserving only the strongest signals. In a Bayesian setting, sparsity can be encouraged through prior specification. Spike-and-slab priors \citep{mitchell1988bayesian, george1993variable} invoke a mixture of two distributions on $\boldsymbol{\beta}$, one shrinking unimportant coefficients to zero (the spike) and the other allowing important coefficients to take non-zero values on a plausible range (the slab). \citet{george1993variable} introduced the latent binary vector $\boldsymbol{\gamma} \in \{0,1\}^p$ to represent models, leading to the Gaussian-mixture formulation:
\begin{equation}
    \label{eq:spike_slab}
    \beta_j|\gamma_j \sim (1-\gamma_j)N(0, \tau^2_j) + \gamma_jN(0, c_j^2\tau_j^2)
\end{equation}
for some suitably small $\tau_j$ and suitably large $c_j$. Priors on $\boldsymbol{\gamma}$ such as  $p(\boldsymbol{\gamma}) = w_{|\boldsymbol{\gamma}|} \binom{p}{|\boldsymbol{\gamma}|}^{-1}$, where $w_{|\boldsymbol{\gamma}|}$ is a weight on the model size, can further encourage sparsity by placing higher probability mass on simpler models i.e., $\boldsymbol{\gamma}$ vectors with fewer elements equal to one. \\

For BVS, the original Stochastic Search Variable Selection (SSVS) constructed a Gibbs Sampler targeting $\boldsymbol{\gamma}$ values with high posterior probability. However, continuous analogues of \eqref{eq:spike_slab} have since been developed to facilitate more sophisticated MCMC algorithms such as No U-Turn Sampling (NUTS) \citep{cont_spike_slab}. For example, MCMC can be used to return draws of the transformed parameters $\boldsymbol{\beta}^{\text{slab}} = \boldsymbol{\beta} \text{ $\cdot$ }  \text{logit}^{-1}(\boldsymbol{\gamma})$, given the prior specification:
\begin{equation}
    \begin{aligned}
    & \boldsymbol{\gamma} \sim \mathcal{N}(\text{logit}(\boldsymbol{\theta}), 1). \\
    & \boldsymbol{\theta} \sim \text{Beta}(b_1, b_2). \\
    & \boldsymbol{\beta} \sim \mathcal{N}(\boldsymbol{0}_p, \tau^2 \mathbb{I}_p). 
    \end{aligned}
\end{equation}
Such an approach easily facilitates variable selection, by computing credible intervals from the draws of $\boldsymbol{\beta}^{\text{slab}}$ and returning variables whose intervals do not contain zero.
\subsection{Gaussian Process Regression}
\label{sec:gps}
Gaussian Process (GP) regression \citep{Rasmussen2006Gaussian} offers a flexible, non-parametric approach to modelling the relationship between input-output pairs, and can capture non-linear dependencies that are not well represented by standard regression models. Let $\mathcal{X} = [\boldsymbol{x}_1,\dots , \boldsymbol{x}_N]^T \in \mathbb{R}^{N\times P}$ and $\boldsymbol{y} \in \mathbb{R}^N$ denote generic inputs and outputs respectively, distinct from the aforementioned explanatory covariates $\textbf{X} \in \mathbb{R}^{n \times p}$ and response vector $\textbf{y} \in \mathbb{R}^n$. \\

It is typically assumed that, for $i=1,\dots , N$, $y_i$ is a noisy observation from a latent function $f$ such that $y_i = f(\boldsymbol{x}_i) + e_i$, where $e_i \sim \mathcal{N}(0, \varsigma^2)$ is Gaussian noise. A Gaussian process prior is then placed on $f$, such that
\begin{equation}
\label{eq:gp}
    f(\boldsymbol{x}) \sim \mathcal{GP}\left(m(\boldsymbol{x}), k(\boldsymbol{x}, \boldsymbol{x}')\right)
\end{equation}
which defines a joint distribution over function values. In particular, for any finite collection of input locations, the corresponding function values follow a multivariate Normal distribution. In \eqref{eq:gp}, $m(\boldsymbol{x})$ represents the mean function, often taken to be zero, and $k(\boldsymbol{x}, \boldsymbol{x}')$ is the kernel which controls the correlation between the function values of two corresponding inputs $\boldsymbol{x}$ and $\boldsymbol{x}'$. For observed data $(\mathcal{X}, \boldsymbol{y})$ and future inputs $\mathcal{X}^*$ for which the output is to be predicted, the joint distribution of function values at observed and future cases is Gaussian. This implies that the predictive distribution over function values at $\mathcal{X}^*$, conditioned on the observed data, is also Gaussian. Specifically,
\begin{equation}
    f(\mathcal{X}^*) \,|\, \mathcal{X}, \boldsymbol{y} \sim \mathcal{N} \left( \mu(\mathcal{X}^*), \Sigma(\mathcal{X}^*) \right),
\end{equation}
with predictive mean and covariance given by
\begin{align*}
\mu(\mathcal{X}^*) &= \textbf{K}_{*}^\top (\textbf{K} + \varsigma^2 \textbf{I})^{-1} \boldsymbol{y}, \\
\Sigma(\mathcal{X}^*) &= \textbf{K}_{**} - \textbf{K}_{*}^\top (\textbf{K} + \varsigma^2 \textbf{I})^{-1} \textbf{K}_{*}
\end{align*}
where \textbf{K} is the covariance matrix for the training inputs, $\textbf{K}_{*}$ is the covariance between training and test inputs, and $\textbf{K}_{**}$ is the covariance matrix for the test inputs. The predictive mean corresponds to the posterior expectation of the latent function at the test inputs, expressed as a linear transformation of the observed responses. The predictive covariance quantifies the reduction in prior uncertainty obtained by conditioning on the observed data. \\

GPs are popular tools due to their ability to provide flexible predictions together with principled uncertainty quantification. However, their main limitation is their computational cost, as fitting a GP scales with approximately $\mathcal{O}(N^3)$ due to the inversion of the $N \times N$ covariance matrix. Various forms of sparse GPs have been proposed to combat this \citep{sparse_gps}, but these do not necessarily scale to large $N$. \\

While GP regression is not inherently a tool for variable selection, Section \ref{sec:gp_hmc} will present how GPs can be used in specifying the inputs as MCMC samples and the outputs as the corresponding log posterior densities of those samples using the method of \citet{nemeth2018merging}. This modelling allows for indirect variable selection, by extending the limited information from the samples into posterior estimation over the entire support of the regression coefficients.

\section{Divide-and-Conquer Methods}
\label{sec:background_methods}
Distributing the computational workload across multiple computer cores, particularly within an embarrassingly parallel framework \citep{neiswanger2014}, can substantially reduce computation time. This section presents a BVS-focused overview of current approaches relevant to the proposed framework that utilise parallelisation by splitting data into subsets and distributing these across cores. The current literature on splitting data can be broadly divided into two categories: splitting on $p$ and splitting on $n$, both of which are discussed here.

\subsection{Split by $p$}
\label{sec:split_p}
Data possessing a large number of covariates, particularly when $p \gg n$, present numerous problems for accurate variable selection such as non-identifiability, overfitting, and multicollinearity. To tackle these issues, \citet{liang_2015} introduced the split-and-merge (SAM) approach for high-dimensional Bayesian variable selection, which consists of two steps. First, the original high dimensional dataset is split into low dimensional subsets via columns. BVS is then performed at significance level $\alpha_1$ on each subset in parallel and the selected variables from each subset are aggregated together. Finally, BVS is performed at significance level $\alpha_2$ on the aggregated set to return a final set of selected variables. The authors recommend taking $\alpha_1 > \alpha_2$ to reduce the number of important variables being dropped while working with the smaller subsets, but mention that the chosen $\alpha$ values are not crucial to the performance of the algorithm. The procedure is consistent a-posteriori under mild conditions and computational efficiency is aided by its embarrassingly parallel structure. A more detailed account of the procedure is given in Algorithm \ref{alg:sam} and in \citet{liang_2015}. While SAM can reduce the computational challenges associated with large $p$, it remains inherently reliant on the BVS methods applied to each subset, and therefore still scales poorly, particularly as $n$ increases. Although this can be addressed using LASSO based approaches such as that of \citet{basil}, doing so sacrifices some of the advantages that BVS methods offer for variable selection compared with classical penalised shrinkage methods.

\begin{algorithm}[t]
\SetAlgoLined
\textbf{Input:}
     Response vector \textbf{y},
     design matrix, \textbf{X},
     binary indicator vector $\boldsymbol{\gamma}$,
     significance level 1 $\alpha_1$,
     significance level 2 $\alpha_2$,
     number of subsets $K$.

\textbf{Algorithm:}
\begin{enumerate}
    \item Partition \textbf{X} and $\boldsymbol{\gamma}$ into $K$ non-overlapping subsets $\textbf{X}_{\cdot,1},\dots,\textbf{X}_{\cdot,K}$ and $\boldsymbol{\gamma}_1,\dots,\boldsymbol{\gamma}_K$ where $\textbf{X}_{\cdot,k}$ is a design matrix consisting of all the rows in $\textbf{X}$ but only the columns corresponding to the indices in $\boldsymbol{\gamma}_k$.
    \item For $(k=1,\ldots,K)$:
    \begin{enumerate}
        \item Perform BVS on $\boldsymbol{\gamma}_k$ at significance level $\alpha_1$, based on the posterior $p(\boldsymbol{\gamma}_k|\textbf{y}, \textbf{X}_{\cdot,k})$.
    \item Denote the set of predictors selected in $\boldsymbol{\gamma}_{k}$ by $\boldsymbol{\tilde{\gamma}}_{k}$.
    \end{enumerate}
    \item Form $\boldsymbol{\tilde{\gamma}} = \bigcup_{k=1}^K \boldsymbol{\tilde{\gamma}}_k$, and perform BVS at significance level $\alpha_2$, using the posterior $p(\boldsymbol{\tilde{\gamma}}|\textbf{y}, \tilde{\textbf{X}})$.
\end{enumerate}
\textbf{Output:}
     Final set of selected variables $\boldsymbol{\gamma}^* \subseteq \boldsymbol{\gamma}$.
\caption{Adapted from Split-and-Merge (SAM) Algorithm \citep{liang_2015}}
\label{alg:sam}
\end{algorithm}

\subsection{Split by $n$}
\label{sec:split_n}

Methods which partition datasets by rows aim to address the principal computational bottleneck induced by large $n$: the expensive evaluation of the likelihood $p(\textbf{y}|\boldsymbol{\beta}, \textbf{X})$. The dataset is split into $L$ batches, each associated with a sub posterior
\begin{equation}
\label{eqn:subpost}
    p(\boldsymbol{\beta}|\textbf{y}_l, \textbf{X}_l) \propto p(\textbf{y}_l|\boldsymbol{\beta}, \textbf{X}_l)p(\boldsymbol{\beta})^{\frac{1}{L}}
\end{equation}
where the fractionated prior ensures that prior information is not over represented when the sub posterior contributions are later recombined. Each core is then assigned the task of generating MCMC draws from one of the sub posterior distributions. The principal challenge is to recover inference on the full posterior from these batch level draws, which is typically achieved by assuming a product of experts decomposition,
\begin{equation}
    p(\boldsymbol{\beta}|\textbf{y}, \textbf{X}) \propto \prod_{l=1}^L p(\boldsymbol{\beta}|\textbf{y}_l, \textbf{X}_{l}) \propto p(\boldsymbol{\beta}) \prod_{l=1}^L  p(\textbf{y}_l|\boldsymbol{\beta}, \textbf{X}_l)
\end{equation}
This section provides a review of current strategies to recombine the sub-posterior draws to arrive at a consensus belief on the full posterior.

\subsubsection{Consensus Monte Carlo (CMC)}
\label{sec:cmc}
To address the task of recombining sub-posterior draws, \citet{consensus} presented the Consensus Monte Carlo (CMC) framework, which tackles the problem through weighted averaging. Let $\boldsymbol{\beta}_l^{(j)}$ denote the $j^{\text{th}}$ draw from the $l^{\text{th}}$ sub-posterior and let $\boldsymbol{\beta}_{\text{con}}$ denote the $J \times p$ matrix of consensus draws, then the $j^{\text{th}}$ row of $\boldsymbol{\beta}_{\text{con}}$ is computed as:
\begin{equation}
    \boldsymbol{\beta}_{\text{con}}^{(j)} = \left(\sum_{l=1}^L W_l \right)^{-1} \sum_{l=1}^L W_l \boldsymbol{\beta}_l^{(j)}
\end{equation}
where $W_l$ is the weight matrix associated with the $l^{\text{th}}$ sub-posterior, three forms of which are explored in \citet{consensus}. One approach is to use the draws $\boldsymbol{\beta}_l^{(1:J)}$ to compute a Monte Carlo estimate of the covariance of the $l^{\text{th}}$ sub-posterior and set $W_l$ equal to this estimate. For particularly large problems, only the variance term may be retained such that $W_l$ is a diagonal weight matrix. In this work, these strategies are referred to as empirical covariance weighting and diagonal empirical covariance weighting respectively. Finally, if covariances can be safely ignored, $W_l$ can be taken as the identity matrix to maximise computational efficiency.
\\

Three additional methods building on the initial CMC framework are compared in \citet{scott2017comparing}, namely sequential, kernel and mixture consensus Monte Carlo (SCMC, KCMC, MxCMC). SCMC and KCMC respectively use importance resampling and kernel density estimation (KDE) for recombination, but both are limited in that they scale poorly with $p$. While MxCMC copes better in high-dimensional scenarios, it is substantially more computationally expensive, and is presented assuming two cores are used, with extension to more cores being non-trivial. A general drawback of CMC methods is their inability to produce a representative sample from more complex posterior distributions, such as those with multiple modes.

\subsubsection{Gaussian Process-Hamiltonian Monte Carlo (GP-HMC)}
\label{sec:gp_hmc}

One potential limitation of the CMC methods is that recombination relies solely on the MCMC draws from the sub posterior distributions. \citet{nemeth2018merging} proposed to also utilise the corresponding log density values computed at each iteration of the MCMC runs. In their approach, a GP is used to model each log sub posterior distribution, treating the draws as inputs and the associated log density values as outputs. Since the product of experts decomposition can be expressed as a sum on the log scale,
\begin{equation}
\label{eqn:log_subpost}
    \log p(\boldsymbol{\beta}|\textbf{y}, \textbf{X}) \propto \sum_{l=1}^L \log p(\boldsymbol{\beta}|\textbf{y}_l, \textbf{X}_{l})
\end{equation}
the full posterior is modelled by a sum of GPs, which is itself a GP.
To simplify notation, define $\ell (\boldsymbol{\beta})^{(l)} := \log p(\boldsymbol{\beta}|\textbf{y}_l, \textbf{X}_{l})$. A GP prior, $\ell(\boldsymbol{\beta})^{(l)} \sim GP(m_l(\boldsymbol{\beta}), S_l(\boldsymbol{\beta}))$, is used, where $m_l(\cdot)$ and $S_l(\cdot)$ are the prior mean and covariance functions respectively. Like in CMC, each core runs an MCMC algorithm to generate draws from the sub-posteriors $p(\boldsymbol{\beta}|\textbf{y}_l, \textbf{X}_l)$, but this time the log sub-posterior density of these draws are also retained. The MCMC samples and their densities are then used to train the GP model and the log sub-posterior density of new samples $\boldsymbol{\beta}_{\text{new}}$ is modelled as:
\begin{equation}
\label{eqn:gp_approx}    \ell(\boldsymbol{\beta}_{\text{new}})^{(l)}|\boldsymbol{\beta}^{(1:J)}_l, \ell (\boldsymbol{\beta}^{(1:J)}_l ) \sim \mathcal{N} \left( \mu_l(\boldsymbol{\beta_{\text{new}}}), \Sigma_l(\boldsymbol{\beta_{\text{new}}}) \right)
\end{equation}
where $\mu_l(\cdot)$ and $\Sigma_l(\cdot)$ are the posterior mean and covariance functions for the $l^{\text{th}}$ GP, respectively. Combining equations (\ref{eqn:log_subpost}) and (\ref{eqn:gp_approx}) gives the approximation to the full log posterior:
\begin{equation}
\ell(\boldsymbol{\beta}_{\text{new}})|\boldsymbol{\beta}^{(1:J)}, \ell(\boldsymbol{\beta}^{(1:J)}) \sim \mathcal{N} \left( \sum_{l=1}^L \mu_l(\boldsymbol{\beta_{\text{new}}}), 
    \sum_{l=1}^L \Sigma_l(\boldsymbol{\beta_{\text{new}}}) \right).
\end{equation}
To obtain a representative sample, one may instead target the expectation of the full posterior density under the GP approximation:
\begin{equation}
    \hat \pi(\boldsymbol{\beta}) \propto \mathbb{E}[\exp (\ell(\boldsymbol{\beta}_{\text{new}})|\boldsymbol{\beta}^{(1:J)}, \ell(\boldsymbol{\beta}^{(1:J)}))] = \exp \left(\sum_{l=1}^L [\mu_l(\boldsymbol{\beta)} + \frac{1}{2} \Sigma_l(\boldsymbol{\beta})] \right).
\end{equation}
The final set of consensus draws is obtained by implementing an HMC sampler with $\hat \pi(\boldsymbol{\beta})$ as the target distribution. \\

\citet{nemeth2018merging} showed that incorporating the log sub posterior density values within GP–HMC led to more accurate posterior inference for particularly complex distributions, such as warped Gaussian targets and rare Bernoulli events, compared with approaches that rely only on shifting and re weighting the sub posterior samples. However, fitting the GP surrogates substantially increases computational cost, and the authors therefore recommend reserving this approach for problems involving sufficiently complex posterior structure.

\section{The Bi-Directional Splitting Framework}
All of the methods described in Section \ref{sec:background_methods} split the data on either rows or columns. This section presents the novel bi-directional BVS splitting framework that combines the previous methods to allow for efficient variable selection in datasets with a huge number of both rows and columns. This section begins by describing how data are split in Section \ref{sec:phase0}. Next, the two phase merging procedure is discussed in Sections \ref{sec:phase1} and \ref{sec:phase2}. Algorithm \ref{alg:my_algorithm} describes the implementation of the framework used in the computer experiments and real data case study. All associated code can be found in \href{https://github.com/aaron-coats/bi_d_split_algorithm}{this GitHub repository}. \\

\begin{figure}[!h]
\centering
\begin{tikzpicture}[node distance=1.8cm]
        \node (start) [xshift = -2cm] {$p(\boldsymbol{\beta}_k | \textbf{y}_l, \textbf{X}_{l,k})$};
        \node (kgiveny) [below of=start, yshift=-0cm, xshift=0cm] {$p(\boldsymbol{\beta}_k|\textbf{y}, \textbf{X}_{\cdot,k})$};
        \node (bgivenl) [right of=start, yshift=-0cm, xshift=2cm] {$p(\boldsymbol{\beta}|\textbf{y}_l, \textbf{X}_{l,\cdot})$};
        \node (end) [right of=kgiveny, yshift = -0cm, xshift=2cm] {$p(\boldsymbol{\beta}|\textbf{y}, \textbf{X})$};
        
        \draw [arrow] (start) -- (kgiveny);
        \draw [arrow] (start) -- (bgivenl);
        \draw [arrow] (bgivenl) -- (end);
        \draw [arrow] (kgiveny) -- (end);
        
        \node (note1.1) [below left of=start, yshift=0.6cm, xshift=-0.5cm]{Aggregate};

        \node (note1.2) [below left of=start, yshift=0.2cm, xshift=-0.5cm]{observations};

        \node (note2.1) [below right of=bgivenl, yshift=0.6cm, xshift=0.5cm]{Aggregate};

        \node (note2.2) [below right of=bgivenl, yshift=0.2cm, xshift=0.5cm]{observations};

        \node (note3.1) [above right of=start, xshift = 0.6cm, yshift = -0.6cm]{Aggregate variables};

        \node (note4.1) [below right of=kgiveny, xshift = 0.6cm, yshift = 0.6cm]{Aggregate variables};
        
    \end{tikzpicture}
    \caption{Diagram of possible routes to go from the starting distributions $p(\boldsymbol{\beta}_k| \textbf{y}_l, \textbf{X}_{l,k})$ to the goal $p(\boldsymbol{\beta}|\textbf{y}, \textbf{X})$}
\label{fig:steps}
\end{figure}
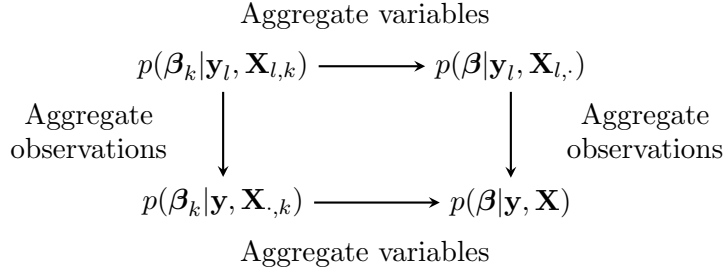

The framework decomposes the task of merging batches split in two directions into two sequential smaller tasks: Merging across columns and merging across rows. Consequently, there are two possible routes from the collection of sub posteriors to the full posterior, as illustrated in Figure \ref{fig:steps}. In this work, column-aggregation is performed first, followed by row-aggregation. This ordering is motivated by the fact that the methods described in Section \ref{sec:split_n} are known to perform poorly in high dimensional settings, and would only need to be used in the second stage. Performing an initial variable screening step therefore helps address the large $p$ setting while also making the subsequent large $n$ problem more tractable.

\label{sec:our_method}

\subsection{Phase 0: Splitting the Data}
\label{sec:phase0}
Consider the sets $\mathcal{N} = \{1,\dots, n\}$ and $\mathcal{P} = \{1,\dots,p\}$ containing the row and column indices respectively. First, $\mathcal{N}$ and $\mathcal{P}$ are respectively partitioned into $L$ and $K$ non-overlapping, approximately equal-sized subsets $\{\mathcal{N}_1, \dots , \mathcal{N}_L \}$ and $\{\mathcal{P}_1, \dots , \mathcal{P}_K \}$ where each $\mathcal{N}_l \subset \mathcal{N}$ and $\mathcal{P}_k \subset \mathcal{P}$. The partitioning satisfies
\begin{equation}
\begin{aligned}
    \mathcal{N}_l \cap \mathcal{N}_{l'} = \emptyset ~\forall~ l \neq l' & \qquad \mathcal{P}_k \cap \mathcal{P}_{k'} = \emptyset ~\forall~ k \neq k' \\ 
    \bigcup_{l=1}^L \mathcal{N}_l = \mathcal{N}, & \qquad \bigcup_{k=1}^K \mathcal{P}_k = \mathcal{P} \\ \left\lfloor\frac{n}{L}\right\rfloor \leq |\mathcal{N}_l| \leq \left\lceil \frac{n}{L} \right\rceil, & \qquad \left\lfloor \frac{p}{K} \right\rfloor \leq |\mathcal{P}_k| \leq \left\lceil \frac{p}{K} \right\rceil
    \end{aligned}
\end{equation}
 $\forall$ $l=1,\dots,L$ and $k=1,\dots K$. This collection of subsets is then used to split the two data objects, $\textbf{y} \in \mathbb{R}^n$ and $\textbf{X} \in \mathbb{R}^{n\times p}$, into $L \cdot K$ batches. \\

Batch $(l,k)$ will contain two objects. The first is the sub-response vector $\textbf{y}_l \in \mathbb{R}^{|\mathcal{N}_l|}$ which is formed by extracting the elements of \textbf{y} corresponding to the indices in $\mathcal{N}_l$. The second is the sub-design matrix $\textbf{X}_{l,k} \in \mathbb{R}^{|\mathcal{N}_l| \times |\mathcal{P}_k|}$ which is formed by extracting the rows and columns of \textbf{X} corresponding to the indices in $\mathcal{N}_l$ and $\mathcal{P}_k$, respectively. It may be beneficial, where possible, to allocate highly correlated columns of $\textbf{X}$ to separate sub matrices in order to mitigate potential adverse effects of multicollinearity. Furthermore, values of $L$ and $K$ such that each sub-matrix has more rows than columns are preferable, to avoid problems associated with $p>n$. Splitting the data in this way leads to the sub-posterior
\begin{equation}
    p(\boldsymbol{\beta}_k|\textbf{y}_l, \textbf{X}_{l,k}) \propto p(\textbf{y}_l|\boldsymbol{\beta}_k, \textbf{X}_{l,k}) p(\boldsymbol{\beta}_k)^{\frac{1}{L}}
\end{equation}
where $\boldsymbol{\beta}_k \in \mathbb{R}^{|\mathcal{P}_k|}$ is the vector of regression coefficients corresponding to the columns of each $\textbf{X}_{l,k}$, $\forall$ $l=1,\dots,L$. \\

It can be seen that each $\boldsymbol{\beta}_k$ is estimated $L$ times, since it is associated with $L$ different sub-matrices. Consequently, the elements of $\boldsymbol{\beta}_k$ are fixed across the $L$ analyses, meaning that none of the coefficients in $\boldsymbol{\beta}_k$ are ever analysed alongside coefficients of $\boldsymbol{\beta}_{k'}$ $\forall$ $k \neq k'$, and therefore potentially useful joint-covariate information could be lost. For this reason, an alternative partitioning scheme is also proposed, wherein the subsets $\{\mathcal{P}_{k=1}^{(l=1)}, \dots \mathcal{P}_{k=K}^{(l=1)}, \dots \mathcal{P}_{k=1}^{(l=L)}, \dots \mathcal{P}_{k=K}^{(l=L)} \}$ are formed, satisfying
\begin{equation}
    \mathcal{P}_{k}^{(l)} \cap \mathcal{P}_{k'}^{(l)} = \emptyset ~\forall~ k\neq k', \quad \bigcup_{k=1}^K \mathcal{P}_k^{(l)} = \mathcal{P}, \quad \left|\mathcal{P}_k^{(l)}\right| \approx \frac{p}{K} \quad \forall~ l=1,\dots,L
\end{equation}

The sub-matrix $\textbf{X}_{l,k}$ is now formed via the indices in $\mathcal{N}_l$ and $\mathcal{P}_k^{(l)}$. At present, for every $l$, the indices $\{1, \dots , p\}$ are randomly shuffled before being partitioned into subsets. This technique will be referred to as `covariate shuffling', but more statistically principled partitioning strategies remain an interesting area of future work. The described splitting methods are illustrated through a simple example with $n = 4$, $p = 4$, $L = 2$, and $K = 2$, shown in \eqref{eqn:split_example}.

\begin{equation}
\label{eqn:split_example}
\begin{aligned}
&\textrm{\textbf{Original}} \\
    &\textbf{y} = \begin{bmatrix}
           \text{y}_1 \\
           \text{y}_2 \\
           \text{y}_3 \\
           \text{y}_4
         \end{bmatrix}, \hspace{0.75cm}
         \boldsymbol{\beta} = \begin{bmatrix}
           \beta_1 \\
           \beta_2 \\
           \beta_3 \\
           \beta_4
         \end{bmatrix}, \hspace{0.75cm}
     \textbf{X} = \begin{bmatrix}
        x_{1,1} & x_{1,2} & x_{1,3} & x_{1,4} \\
        x_{2,1} & x_{2,2} & x_{2,3} & x_{2,4} \\
        x_{3,1} & x_{3,2} & x_{3,3} & x_{3,4} \\
        x_{4,1} & x_{4,2} & x_{4,3} & x_{4,4}
    \end{bmatrix} \\ &  \\
    & \textrm{\textbf{Simple Split}} \\
    & \textbf{y}_1 = \begin{bmatrix}
            \text{y}_1 \\ \text{y}_2
        \end{bmatrix}, \hspace{0.5cm} 
        \boldsymbol{\beta}_1 = \begin{bmatrix}
            \beta_1 \\ \beta_2
        \end{bmatrix} \hspace{.5cm}
        \textbf{X}_{1,1} = \begin{bmatrix}
            x_{1,1} & x_{1,2} \\ x_{2,1} & x_{2,2}
        \end{bmatrix}, \hspace{0.5cm}
        \textbf{X}_{1,2} = \begin{bmatrix}
            x_{1,3} & x_{1,4} \\
            x_{2,3} & x_{2,4}
        \end{bmatrix}
        \\
        & \textbf{y}_2 = \begin{bmatrix}
            \text{y}_3 \\ \text{y}_4
        \end{bmatrix}, \hspace{0.5cm}
        \boldsymbol{\beta}_2 = \begin{bmatrix}
            \beta_3 \\ \beta_4
        \end{bmatrix} \hspace{.5cm}
        \textbf{X}_{2,1} = \begin{bmatrix}
            x_{3,1} & x_{3,2} \\ x_{4,1} & x_{4,2}
        \end{bmatrix}, \hspace{0.5cm}
        \textbf{X}_{2,2} = \begin{bmatrix}
            x_{3,3} & x_{3,4} \\
            x_{4,3} & x_{4,4}
        \end{bmatrix} \\ & \\
        & \textrm{\textbf{Covariate Shuffling}} \\
        & \textbf{y}_1 = \begin{bmatrix}
            \text{y}_1 \\ \text{y}_2
        \end{bmatrix}, \hspace{0.5cm}
        \boldsymbol{\beta}_1^{(1)} = \begin{bmatrix}
            \beta_1 \\ \beta_2
        \end{bmatrix} \hspace{.5cm}
        \boldsymbol{\beta}_1^{(2)} = \begin{bmatrix}
            \beta_1 \\ \beta_3
        \end{bmatrix}, \hspace{.5cm} 
        \textbf{X}_{1,1} = \begin{bmatrix}
            x_{1,1} & x_{1,2} \\ x_{2,1} & x_{2,2}
        \end{bmatrix}, \hspace{0.5cm}
        \textbf{X}_{1,2} = \begin{bmatrix}
            x_{1,3} & x_{1,4} \\
            x_{2,3} & x_{2,4}
        \end{bmatrix}
        \\
        & \textbf{y}_2 = \begin{bmatrix}
            \text{y}_3 \\ \text{y}_4
        \end{bmatrix}, \hspace{0.5cm}
        \boldsymbol{\beta}_2^{(1)} = \begin{bmatrix}
            \beta_3 \\ \beta_4
        \end{bmatrix}  \hspace{.5cm}
        \boldsymbol{\beta}_2^{(2)} = \begin{bmatrix}
            \beta_2 \\ \beta_4
        \end{bmatrix}, \hspace{.5cm} 
        \textbf{X}_{2,1} = \begin{bmatrix}
            x_{3,1} & x_{3,3} \\ x_{4,1} & x_{4,3}
        \end{bmatrix}, \hspace{0.5cm}
        \textbf{X}_{2,2} = \begin{bmatrix}
            x_{3,2} & x_{3,4} \\
            x_{4,2} & x_{4,4}
        \end{bmatrix}
         \end{aligned}
\end{equation}

\subsection{Phase One: Merging Columns}
\label{sec:phase1}
\label{sec:phase_one_theory}
After the data are split, core $(l,k)$ receives the objects $\textbf{y}_l$ and $\textbf{X}_{l,k}$ and performs a standard BVS technique to return the set of indices $I_l^{(k)} \subset \mathcal{P}$ which represents the columns of $\textbf{X}_{l,k}$ that are considered important in modelling $\textbf{y}_l$. Next, for every $l$, the $K$ workers identified by the index pairs $(l,k)_{k=1}^K$ aggregate their indices to form the sets $I_l = \bigcup_{k=1}^K I_l^{(k)}$. A final round of BVS is then performed on $I_l$ to return the set $I^*_l \subseteq I_l$. This is essentially equivalent to independently performing $L$ runs of the SAM method \citep{liang_2015} with the maximum iteration value $T$ set to 1. \\

 Since each variable is processed $L$ times, there are $L$ opportunities for it to be selected. The methods for merging across rows require a fixed set of variables, and since it is not guaranteed that $I^*_l = I^*_{l'}$ $\forall$ $l \neq l'$, an agreement must be made as to which variables should be sent to the second phase. To do this, the proportion $p_i = \frac{1}{L}\sum_{l=1}^L \mathbb{I}\{ i \in I_l^*\}$ is computed for every index $i \in \bigcup_{l=1}^LI_l^*$, and the set $I^* = \{i: p_i \geq r \}$ is returned, where $r \in [0,1]$ is a user-specified threshold. The subset $I^*$ induces a corresponding parameter vector $\boldsymbol{\beta}^* \in \mathbb{R}^{|I^*|}$, corresponding to the variables in $I^*$. Information about $\boldsymbol{\beta}^*$ must now be merged across observations.

 \begin{algorithm}[!t]
\SetAlgoLined
\textbf{Input:}
     Response vector \textbf{y}, design matrix \textbf{X}, number of row-splits $L$, number of column splits $K$, latent inclusion vector $\boldsymbol{\gamma}$, prior inclusion probability vector $\boldsymbol{\theta}$, prior inclusion hyperparameters $\boldsymbol{b} = (b_1, b_2)$, significance levels $\alpha_1$ and $\alpha_2$, selection threshold $r$.
\\ \vspace{0.3cm}
\textbf{Algorithm:}
\begin{enumerate}
    \item Partition \textbf{y} and \textbf{X} respectively into the sub-vectors $\textbf{y}_l$ and sub-matrices $\textbf{X}_{l,k}$ for $l=1,\dots,L$ and $k=1,\dots,K$. Denote by $\boldsymbol{\beta}_k^{(l)}$ and $\boldsymbol{\gamma}_k^{(l)}$ respectively, the coefficient and inclusion vectors corresponding to $\textbf{X}_{l,k}$.

    \item For $l=1,\ldots,L$:
    \begin{enumerate}
        \item For $k=1,\ldots,K$:
    \begin{enumerate}
        \item Perform BVS on the sub-posterior $p(\boldsymbol{\beta}_k^{(l)}|\textbf{y}_l, \textbf{X}_{l,k}, \boldsymbol{\gamma}_k^{(l)}, \boldsymbol{\theta}, \boldsymbol{b})$ at significance level $\alpha_1$.
        \item Denote the set of variables selected by $I_{l}^{(k)}$.
    \end{enumerate}
    \item Form $I_l = \bigcup_{k=1}^K I_{l}^{(k)}$.
    \item Perform BVS at significance level $\alpha_2$ on the sub-posterior $p(\boldsymbol{\beta}^{(l)} | \textbf{y}_l, \textbf{X}^{(l)}, \boldsymbol{\gamma}^{(l)}, \boldsymbol{\theta}, \boldsymbol{b})$, where ${\boldsymbol{\beta}}^{(l)}, \textbf{X}^{(l)}$ and $\boldsymbol{\gamma}^{(l)}$ represent the variables corresponding to the indices in $I_l$.
    \item Denote by $I_l^*$ all the variables selected after the second round.
    \end{enumerate}
    \item For every variable index $i$ in $\bigcup_{l=1}^L I_l^*$, compute the proportion $p_i = \frac{1}{L} \sum_{l=1}^L \mathbb{I}\{i \in I_l^* \}$.
    \item Set $I^* = \{i:p_i \geq r \}$ and define $\boldsymbol{\beta}^* \subset \boldsymbol{\beta}$ as the coefficient vector corresponding to the indices in $I^*$.

    \item For $l=1,\ldots ,L$: \\ \hspace{0.1cm} Generate $J$ MCMC draws, $\left\{\boldsymbol{\beta}^{*(j)}_l \right\}_{j=1}^J$, from the sub-posterior $\boldsymbol{\beta}^*|\textbf{y}_l, \textbf{X}_{l,*}$. If using GP-HMC for row-merging,
    also save the log sub-posterior densities $\ell \left(\boldsymbol{\beta}^{*(j)}_l\right) := \log p(\boldsymbol{\beta}^{*(j)}|\textbf{y}_l, \textbf{X}_{l,*})$.
    \item Merge the $L$ draws matrices into one matrix of consensus draws.
    \item Compute 95\% credible intervals for $\boldsymbol{\beta}^*$ using the consensus draws, and return all variables whose intervals do not contain zero.
\end{enumerate}
\textbf{Output:}
     Final set of selected variables.
\caption{The Bi-Directional Splitting Algorithm}
\label{alg:my_algorithm}
\end{algorithm}

\subsection{Phase Two: Merging Rows}
\label{sec:phase2}
\label{sec:phase_two_theory}
The objective is now to address the large $n$ problem by aggregating information about $\boldsymbol{\beta}^*$ across $L$. First, $J$ MCMC draws are generated from the distributions $p(\boldsymbol{\beta}^*|\textbf{y}_l, \textbf{X}_{l,*}) ~\forall~ l=1, \dots , L$, giving draws $\left\{ \left\{\beta^{*(j)} | \textbf{y}_l, \textbf{X}_{l,*} \right\}_{j=1}^J \right\}_{l=1}^L$. These $L$ $J \times |\boldsymbol{\beta}^*|$ matrices of MCMC draws are now to be merged into a single matrix of consensus draws. The methods introduced in \citet{scott2017comparing} and \citet{nemeth2018merging} are useful tools for tackling such a problem. Computationally, the merging at this stage is made a lot cheaper, after implementing the screening procedure in the first phase to massively reduce the dimension of the parameter space. The chosen row-aggregation method produces an approximate draw from the posterior distribution $p(\boldsymbol{\beta}^*|\textbf{y}, \textbf{X}_{\cdot,*})$. Finally, $\forall$ $i \in I^*$, 95\% credible intervals based on the merged draws are computed and all variables whose intervals do not contain zero are returned as the selected variables.   

\section{Experiments}
\label{sec:simulations}
The performance of the proposed framework was assessed with synthetic data in a series of computer experiments. Section \ref{sec:training} describes how small datasets were used to better understand hyperparameter sensitivity and to provide guidelines on hyperparameter choice. The algorithm was then applied to large datasets in Section \ref{sec:testing}, which also compares covariate splitting techniques and consensus methods for the second stage. \\

Variable selection can be considered as a binary classification problem as variables are to be classified as being important or not. Since it was known ahead of time in the experiments which variables truly were important, binary evaluation metrics were used to assess performance. Denoting important variables as `positive' cases and unimportant variables as `negative' cases, the $2\times 2$ confusion matrix is given by
\[
\begin{array}{c|c c}
    & \text{Predicted Positive} & \text{Predicted Negative}
    \\
    \hline
    \text{Actual Positive} & \text{TP} & \text{FN} \\
    \text{Actual Negative} & \text{FP} & \text{TN} \\
\end{array}
\]
where TP (True Positives) and TN (True Negatives) correspond respectively to the number of actual positive and actual negative variables that were correctly classified as such. FP (False Positives) and FN (False Negatives) are the numbers of variables that were incorrectly classified as being positive and negative respectively. As variable selection is more concerned with prediction of the positive class, performance was measured using the $F_1$ score
\begin{equation}
        F_1 = \frac{2 \cdot TP}{2 \cdot TP + FP + FN}
\end{equation}
which may also be interpreted as the harmonic mean of precision (proportion of all correctly identified predicted positives) and recall (proportion of all correctly identified true positives).

\subsection{Hyperparameter Sensitivity \& Guidelines}
\label{sec:training}
This section describes testing of the hyperparameters $L$, $K$, $\alpha_1$, $\alpha_2$, $b_1$, $b_2$, and $r$. Section \ref{sec:setup} describes the generation of the synthetic data as well as the hyperparameter values that were considered. Section \ref{sec:LK_r_training} reports the results for $L$, $K$ and $r$, i.e., the new parameters introduced in this work. As the objective of this stage was only to identify appropriate hyperparameter values, only the CMC identity averaging method was used in phase two due to its computational speed. Comparisons of phase two merging techniques were reserved for Section \ref{sec:testing}.

\begin{figure}[p]
    \centering

    \begin{subfigure}{0.9\linewidth}
        \centering
        \includegraphics[width=\linewidth]{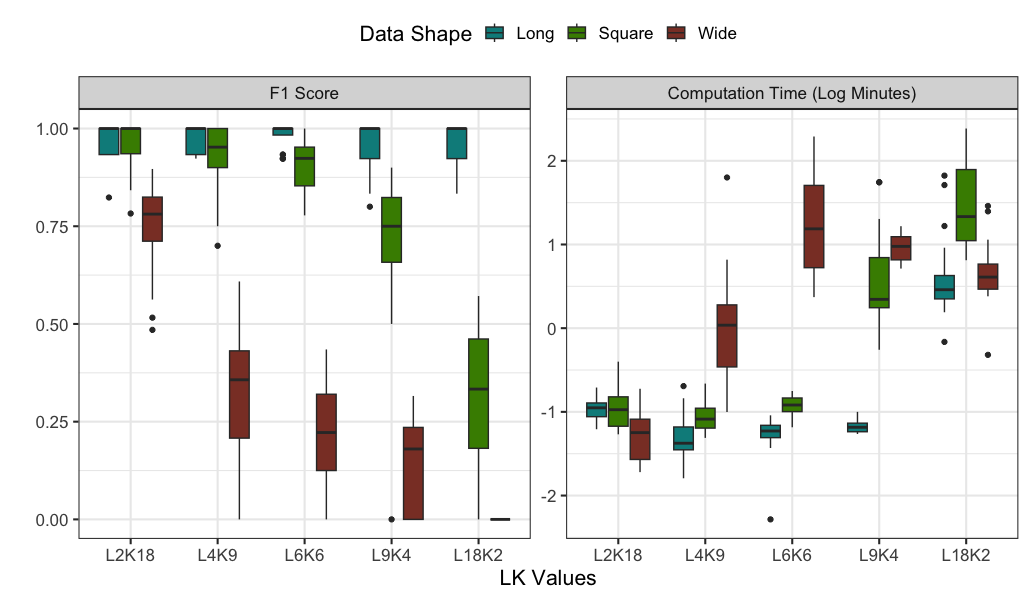}
        \caption{LK Values}
        \label{fig:LK_plot}
    \end{subfigure}

    \vspace{0.5cm}

    \begin{subfigure}{0.9\linewidth}
        \centering
        \includegraphics[width=\linewidth]{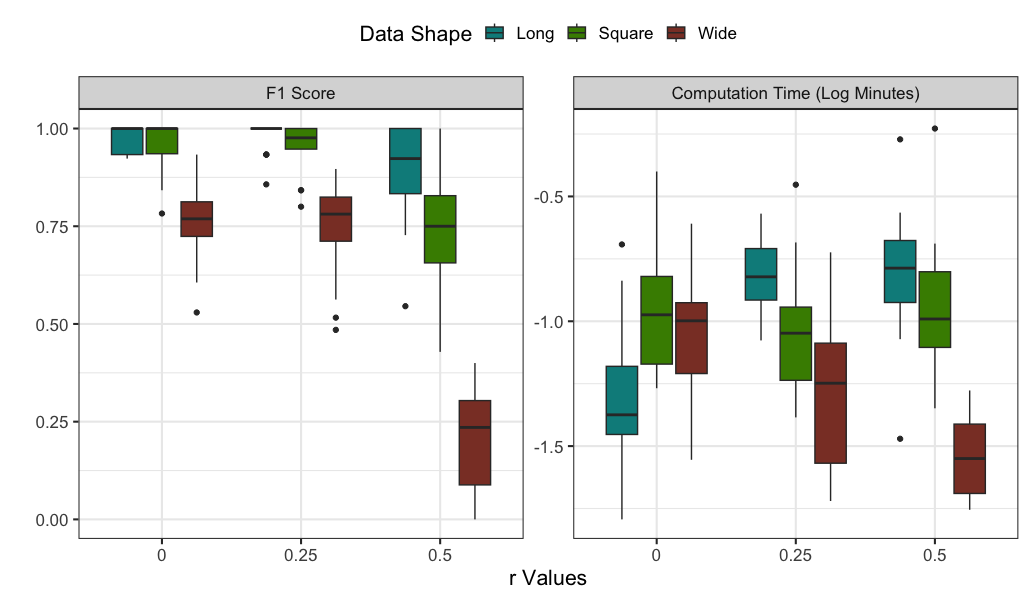}
        \caption{r Values}
        \label{fig:r_plot}
    \end{subfigure}

    \caption{Box-plots showing the distribution of $F_1$ score and computation time for different hyperparameter configurations across each data shape. Figure (a) shows results for all considered values of $L$ and $K$, filtered for the best-performing values of $r$, $\alpha_1$, $\alpha_2$, $b_1$ and $b_2$. Figure (b) shows results for all considered values of $r$, filtered for the best-performing values of $L$, $K$, $\alpha_1$, $\alpha_2$, $b_1$ and $b_2$. Each box consists of 20 points, corresponding to the 20 repetitions of the experiment associated with the relevant hyperparameter configuration.}
    \label{fig:training_results}
\end{figure}

\subsubsection{Setup}
\label{sec:setup}
Data generation involved generating each row of the design matrix \textbf{X} as a $N(\boldsymbol{0}_p, \Sigma_{p\times p})$ random variable, where $\Sigma_{p\times p}$ is a Toeplitz correlation matrix with some correlation coefficient $\rho \in [0,1)$, initially set to 0.3. The vector $\boldsymbol{\beta}_{\text{true}}$ was then initialised as a $p$-dimensional vector of zeros, and $q<p$ of its elements were randomly assigned values with magnitude between 2 and 4. The response \textbf{y} was then generated as a $N(\textbf{X} \boldsymbol{\beta}_{\text{true}}, \sigma^2 \mathbb{I}_n)$ random variable, where initially $\sigma^2=1$. As the structure of the original dataset can be expected to influence the results, three data shapes were used: square ($n = 300, p = 300, q = 10$), long ($n = 500, p = 150, q = 7$), and wide ($n = 150, p = 500, q = 15$). For each shape, 20 replicates were generated. The considered values of the hyperparameters were
\begin{itemize}
    \item $(L,K) \in \{(2,18), (4,9), (6,6), (9,4), (18,2) \}$
    \item $(\alpha_1,\alpha_2) \in \{(0.2,0.1), (0.35,0.25), (0.5,0.4), (0.65,0.55) \}$
    \item $(b_1, b_2) \in \{(0.5, 0.5), (1,1), (1,5), (5,5), (1,10), (1,100) \}$
    \item $r \in \{0, 0.25, 0.5\}$
\end{itemize}
and each of the sixty datasets were analysed once for every combination. The values of $L$ and $K$ were chosen under the assumption that there were $L \cdot K=36$ cores available in order to understand how best to partition the data in terms of number of row/column splits. The values of $\alpha_1, \alpha_2$ and $r$ represent varying degrees of conservativeness in the inference and the values of $b_1$ and $b_2$ accommodated varying beliefs about sparsity. For comparison, each dataset was also analysed using the SAM and CMC algorithms, such that the data was only split along one dimension rather than two like in the Bi-Directional Splitting Algorithm. Finally, since the datasets used in the training setup are comparatively small, an analysis with `no splitting' was used to demonstrate the computational gains from splitting, using a spike-and-slab regression at significance level $\alpha=0.05$.

\subsubsection{$L$, $K$ and $r$}
\label{sec:LK_r_training}

Three new hyperparameters were introduced by the proposed Bi-Directional Splitting Algorithm, and therefore understanding how these values affect the performance of the algorithm is important. First, for every combination of data shape, $L$ and $K$, the combination of $r$, $\alpha_1$, $\alpha_2$, $b_1$ and $b_2$ that produced the highest median $F_1$ score was determined. In the event of a tie, the combination that yielded the lowest median computation time was selected. The results were then filtered to only contain these values of $r$, $\alpha_1$, $\alpha_2$, $b_1$ and $b_2$. \\

Figure \ref{fig:LK_plot} shows that $F_1$ score tended to decrease as $L$ increased and $K$ decreased. There are two reasons for this trend: First, as $L$ becomes large, the number of rows per-block becomes smaller, meaning there is less information available to accurately estimate coefficients and perform variable selection. Secondly, batches become wider with growing $L$ and shrinking $K$, eventually leading to non-identifiability in the design matrix when the number of columns exceeds the number of rows in a batch. This trend was most apparent for wide data, with median $F_1$ score dropping from 0.781 at $(L=2,K=18)$ to 0 at $(L=18,K=2)$. Square data's median $F_1$ dropped from 1 to 0.333 across the same range. However, long data was completely resilient to the different splitting schemes with a median $F_1$ of 1 in every scenario. Long data did perform best under equal $L$ and $K$ as this setting had the tightest variance. Identifying a clear trend in computation time is more difficult, although lower values of $L$ generally resulted in shorter run times. This is also likely due to the resulting batches being of higher dimension, thus increasing the time taken to e.g. perform MCMC. \\

After examining $L$ and $K$, a similar procedure was conducted for $r$. For each value of $r$, the results were filtered to retain only the combination of the other hyperparameters yielding the highest median $F_1$ score, with the lowest median log-computation time selected in the event of a tie. From Figure \ref{fig:r_plot}, the results for $r=0$ and $r=0.25$ were broadly similar, with consistently producing higher $F_1$ scores. Long data had better results under $r=0.25$, retaining the median $F_1=1$ from the previous setting, but exhibiting much tighter variance. Modest changes were seen for square and wide data, the former's $F_1$ dropping from 1 to 0.976 and the latter's increasing from 0.769 to 0.781. However, all three shapes experienced a noticeable drop in performance when increasing $r$ to $0.5$. This was most apparent for wide data whose median $F_1$ dropped to 0.235. \\

Log-computation time for wide data decreased with $r$, likely due to stricter thresholds retaining fewer variables, and thus resulting in cheaper computation down the line. Square data's computation time did not appear to vary greatly between $r$ values, but long data's runs actually got longer on average with increasing $r$.

\subsubsection{Comparison with Competing Methods}
\label{sec:training_comparison}
\begin{figure}[t]
    \centering
    \includegraphics[width=0.9\linewidth]{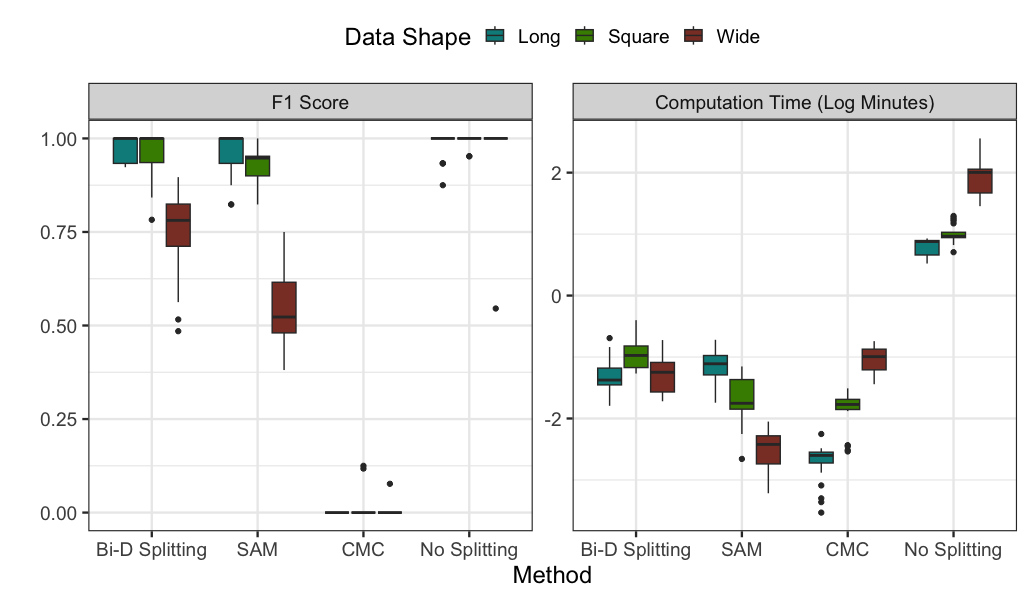}
    \caption{Box-plot showing the distribution of $F_1$ score and log computation time for the proposed framework, as well as the competing one-direction-splitting methods. A `no splitting' analysis, amounting to a spike-and-slab regression with $\alpha=0.05$, was also included to serve as both a sanity check and an illustration of the computational gains made from splitting.}
    \label{fig:methods_comparison}
\end{figure}

As the proposed framework combines methodologies from column-splitting and row-splitting, SAM and CMC were also applied to the datasets in turn, to compare splitting in one versus two directions. Figure \ref{fig:methods_comparison} highlights the advantage of the proposed bi-directional approach over splitting in one direction only. CMC had a median $F_1$ score of 0 for all data shapes, due to its inability to recover any true positive cases in the majority of runs. This is due to the method's inability to cope with high-dimensional problems, further motivating splitting in two directions. SAM was much more competitive with the proposed framework, with a median $F_1$ score of 0.905 compared to 0.933 for bi-directional splitting. SAM was also more efficient than the bi-directional splitting algorithm for square and wide data.

\subsubsection{Guidelines}
To choose fixed values for each parameter to carry forward into the testing stage, the results data were grouped by $L,K,r,\alpha_1, \alpha_2, b_1$ and $b_2$ and the median $F_1$ score and computation time were computed for each combination. The `best' values were then taken to be the row with the highest median $F_1$ score, and the lowest computation time in the event of a tie. Those values are given in Table \ref{tab:trained_params}, and were used on datasets throughout the testing stage.

\begin{table}[b]
    \centering
    \begin{tabular}{c|rrrrrr}
      Data Shape   & $L,K$ & $r$ & $\alpha_1$ & $\alpha_2$ & $b_1$ & $b_2$ \\ \hline 
      Long   & $4, 9$ & 0 & 0.2 & 0.1 & 0.5 & 0.5 \\
      Square & $2, 18$ & 0 & 0.2 & 0.1 & 1 & 5 \\
      Wide & $2,18$ & 0.25 & 0.35 & 0.25 & 5 & 5 \\ 
    \end{tabular}
    \caption{The hyperparameter values from the initial experiments, for each data shape as described in Section \ref{sec:setup}.}
    \label{tab:trained_params}
\end{table}

\subsection{Testing}
\label{sec:testing}
After suitable hyperparameter values were established, the model was used to investigate robustness to data size and to compare phase two consensus methods (Section \ref{sec:big_data_sim}). The model was also used for comparing data splitting techniques in the presence of varying degrees of correlation among the covariates (Section \ref{sec:corr_sim}). 
In the testing stage, datasets were generated in a similar manner to what was described in Section \ref{sec:setup} and any deviations will be outlined in the relevant sections. The two metrics considered throughout were again the $F_1$ score and the computation time.

\subsubsection{Big Datasets \& Phase Two Techniques}
\label{sec:big_data_sim}
\begin{figure}[t]
    \centering
    \includegraphics[width=0.9\linewidth]{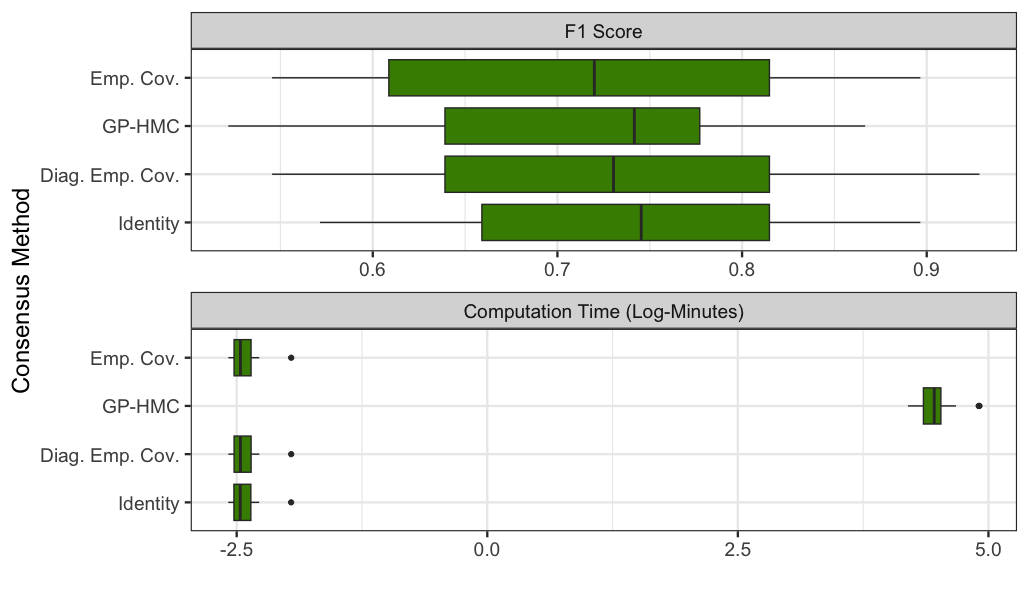}
    \caption{Box-plot showing the distribution of $F_1$ score and log computation time for each phase two approach on the large datasets. Overall, little variation was observed in terms of predictive performance, as can be seen from the substantial overlap in the boxes. Unsurprisingly, the GP-HMC took substantially longer than the three averaging methods.}
    \label{fig:big_data_comparison}
\end{figure}

To investigate the proposed framework's robustness to large datasets, 20 replicates of size $n = 50,000$ and $p = 20,000$ were generated. To establish applicability for more difficult datasets, the residual variance and degree of correlation were increased to $\sigma^2 = 2.5$ and $\rho=0.5$. All 20 datasets underwent identical phase one analysis, using the chosen hyperparameter values from the initial experiments, and sub-posterior draws from each $p(\boldsymbol{\beta}^*|\textbf{y}_l, \textbf{X}_{l,*})$ distribution were generated. These draws were then merged using the three CMC averaging techniques and the GP-HMC method, which are both described in Section \ref{sec:phase_two_theory}. Code for GP-HMC was adapted from \citet{diffusion_gen}. Figure \ref{fig:big_data_comparison} compares the distributions of $F_1$ score and log computation time for the four competing methods. \\

Comparing merging techniques did not find substantial difference in predictive performance among them, as significant overlap in the $F_1$ distributions was present as shown in Figure \ref{fig:big_data_comparison}. The three CMC averaging methods yielded almost identical computation time, whereas GP-HMC was unsurprisingly much slower. However, it should be emphasised here that the GP-HMC algorithm is designed for more complex distributions than those in this work, and experiments in \citet{nemeth2018merging} show the approach excels in sampling from such distributions compared to CMC.

\subsubsection{Correlated Datasets \& Covariate Shuffling}
\label{sec:corr_sim}
Performance of the proposed framework in the presence of varying degrees of correlation in the covariates, as well as potential gains from covariate shuffling \eqref{eqn:split_example} were of interest. The correlation matrix, $\boldsymbol{\Sigma}$, used in the generation of \textbf{X} was given a Toeplitz structure with pre-specified correlation coefficient $\rho$ which can take any real value between 0 and 1. In the experiments, a sequence of $\rho$ values from 0 to 0.9 in increments of 0.1, in addition to a very highly correlated case of 0.99, were considered. For every $\rho$ value, 20 datasets were generated ($n=p=500$) and each replicate was put through the bi-directional splitting algorithm twice, once for each of the two splitting techniques described in Section \ref{sec:phase0}. \\

\begin{figure}[t]
    \centering
    \includegraphics[width=0.9\linewidth]{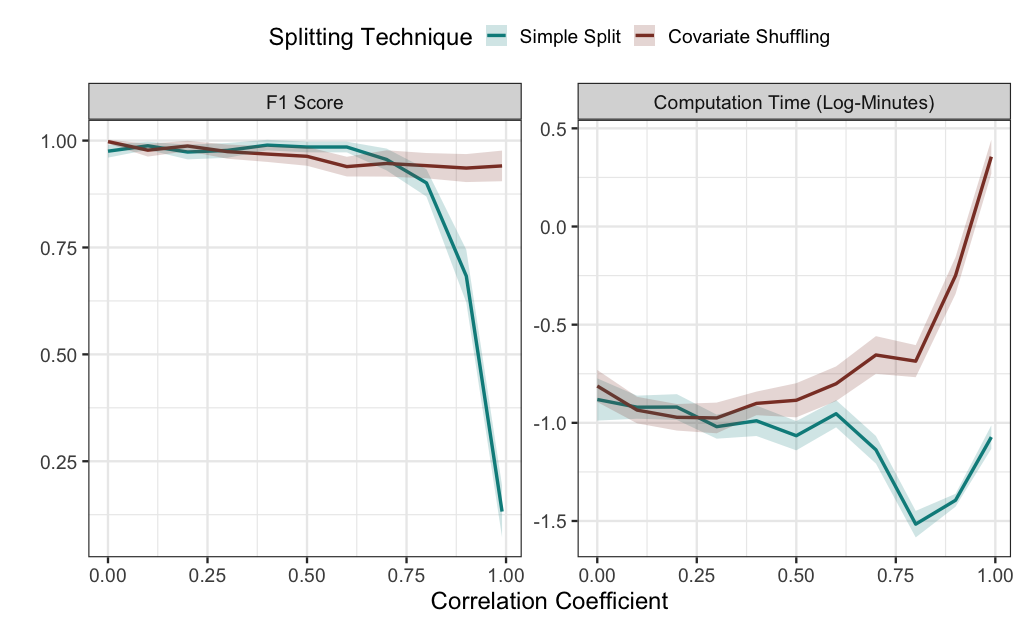}
    \caption{Line-plot showing the distribution of $F_1$ score and log computation time for each correlation coefficient. Simple splitting performs well up until around $\rho=0.6$, beyond which the $F_1$ score drops dramatically. Shuffling, on the other hand, is considerably more robust to increasing correlation, with only a modest drop in $F_1$ score. This does come at the expense of increased computation time as can be seen from the right plot.}
    \label{fig:correlation_comparison}
\end{figure}

Figure \ref{fig:correlation_comparison} shows the distributions of $F_1$ score and log computation time for the varying degrees of correlation and the two splitting techniques. Unsurprisingly, predictive performance deteriorated as correlation grew stronger, but shuffling the covariates did yield more robustness in these settings compared to the simple splitting approach, as can be seen from the higher $F_1$ scores for shuffling. For example, the $F_1$ score for simple splitting dropped from around 1 for lower correlations, to 0.126 for the highly correlated case. Shuffling the covariates yielded a median score of 0.937 in the highly correlated case. This is because shuffling the covariates massively reduced the presence of collinearity in each batch, as highly correlated covariates no longer necessarily appeared together. A trade-off between performance and computation time can be clearly observed, with shuffling the covariates producing much slower runs as correlation grows. This is likely due to a greater number of variables progressing to stage 2 of the algorithm. Although the additional computational cost is arguably justified by the improved performance, the hyperparameters could alternatively be adjusted to achieve computational times closer to those of the simple splitting method.

\section{H3N2 Influenza Data Application}
\label{sec:h3n2}

To demonstrate the Bi-Directional Splitting Algorithm, we consider a dataset of antigenic measurements for influenza A(H3N2) viruses, with the aim of identifying genetic determinants of antigenic drift. The data, which can be found at \href{https://github.com/will-harvey/Flu_g2p_mapping/tree/main/dataset_h3n2}{this GitHub repository}, comprise pairwise measures of antigenic similarity for influenza A(H3N2) viruses obtained using the haemagglutination inhibition (HI) assay \citep{h3n2}. Following its emergence in the 1968 pandemic, H3N2 has been subject to extensive global surveillance to monitor antigenic drift. In an HI assay, antisera raised in ferrets against a reference virus are tested for their ability to inhibit haemagglutination by a test virus. The resulting HI titre, defined as the maximum serum dilution that continues to inhibit haemagglutination, provides a quantitative measure of antigenic similarity between pairs of virus strains. In this work, the response variable is the log-transformed HI titre, measured for 38,751 reference-test virus pairs.


The 616 binary explanatory variables are of two types. The first type corresponds to 219 amino acid site variables, where a value of 1 indicates a difference between the reference and test viruses at a particular position and 0 indicates that there is no difference. Identifying sites associated with antigenic change is the primary objective of the analysis, as these may provide insights into the mechanisms of antigenic drift and inform future vaccine design. The second type comprises 397 phylogenetic variables that capture evolutionary effects associated with the reference and test viruses that are not explained by sequence variation at individual sites. These variables are included as adjustment factors to account for shared evolutionary history and are not themselves the focus of inference.

The dataset was analysed using the bi-directional splitting algorithm with $L=4$ and $K=25$, with the remaining parameters set according to the ``long'' row of Table \ref{tab:trained_params}. The threshold parameter, $r$, was set to 0.4 to encourage a more conservative variable selection procedure, reducing the inclusion of weakly supported variables and improving the interpretability of the resulting model. This was particularly important in the present application, where only a relatively small number of candidate sites could realistically be prioritised for subsequent experimental investigation. The use of the bi-directional splitting algorithm was motivated by the scale of the problem, as previous Bayesian variable selection approaches based on spike-and-slab priors, such as \citet{davies2019improving}, are computationally challenging to apply to datasets of this size.

The proposed method identified 25 sites of interest which are presented with their Maximum A Posteriori (MAP) estimates and 95\% credible interval bounds in Table \ref{tab:sites}. Of these, 21 are located within, or in close proximity to, the established antigenic sites A, B, C, D, and E on the haemagglutinin protein. Previous work by \citet{koel2013substitutions} investigated the genetic basis of major antigenic changes in H3N2 through reverse genetics experiments and identified substitutions responsible for transitions between antigenic clusters; four of the identified sites correspond to substitutions reported in that study. A further 17 have previously been associated with antigenic variation in H3N2 based on the evidence discussed in \citet{h3n2}. In addition, seven are located in or near the receptor-binding site (RBS), the region of the haemagglutinin protein responsible for binding to host cell receptors and a key determinant of viral infectivity and antigenic evolution \citep{wilson1990structural}. The remaining four are not located within recognised antigenic regions and are likely situated towards the base of the haemagglutinin protein. Consequently, these sites have less supporting biological evidence than the other selected sites and should be interpreted with greater caution.

\begin{table}[!t]
    \centering
    \begin{tabular}{c|c|c|c|r|r|r}
        HA Site no. & Antigenic Site & RBS & Koel Site & MAP & 2.5\% & 97.5\% \\ \hline
        
        8 & - &  &  & -.0646 & -.0728 & -.0538 \\
        21 & - &  &  & .1200 & .1103 & .1297 \\
        62 & E &  &  & -.1692 & -.1897 & -.1532 \\
        96 & (E) & ($\checkmark$) &  & .1046 & ..0952 & .1135 \\
        124 & A &  &  & -.1949 & -.2092 & -.1808 \\
        131 & A & ($\checkmark$) &  & -.2414 & -.2527 & -.2312 \\
        144 & A &  &  & -.0783 & -.0897 & -.0659 \\
        145 & A &  & $\checkmark$ & -.1685 & -.1786 & -.1580 \\
        158 & B &  & $\checkmark$ & -.2493 & -.2624 & -.2369 \\
        168 & (B) &  &  & -.0064 & -.0147 & .0023 \\
        169 & (B) &  &  & -.0545 & -.0630 & -.0460 \\
        173 & D &  &  & -.0478 & -.0581 & -.0365 \\
        189 & B & ($\checkmark$) & $\checkmark$ & -.1076 & -.1183 & -.0972 \\
        193 & B & ($\checkmark$) & $\checkmark$ & -.0333 & -.0434 & -.0226 \\
        196 & B & ($\checkmark$) &  & -.1083 & -.1217 & -.0943 \\
        199 & B &  &  & -.0515 & -.0601 & -.0414 \\
        212 & (D) &  &  & -.0626 & -.0781 & -.0477 \\
        213 & (D) &  &  & -.0111 & -.0207 & -.0017 \\
        214 & (D) &  &  & -.1044 & -.1185 & -.0916 \\
        219 & (D) &  &  & .0543 & .0434 & .0644 \\
        225 & (D) & $\checkmark$ &  & -.2822 & -.294 & -.2721 \\
        229 & (D) & ($\checkmark$) &  & .1428 & .1330 & .1525 \\
        262 & E &  &  & -.1701 & -.1815 & -.1586 \\
        309 & - &  &  & -.0642 & -.0737 & -.0537 \\
        326 & - &  &  & -.1245 & -.1353 & -.1157 \\
    \end{tabular}
    \caption{Amino acid sites identified by the bi-directional splitting algorithm, together with their corresponding antigenic site classification, whether the site is located in the RBS, whether the site was identified in \citet{koel2013substitutions}, maximum a posteriori (MAP) coefficient estimates and 95\% credible intervals. Parentheses indicate close proximity.}
    \label{tab:sites}
\end{table}

\section{Discussion}
\label{sec:conclusions}
Performing Bayesian variable selection on modern datasets is computationally difficult, as the number of rows and columns of those datasets grows. Divide-and-Conquer methodology offers attractive gains in speed since the workload can be distributed across multiple computer cores, translating in the BVS context to partitioning the data into subsets and allocating each subset to a core. However, how the data are to be partitioned is not trivial, and current methods only partition by either rows (Section \ref{sec:split_n}) or columns (Section \ref{sec:split_p}). \\

This paper has introduced a novel framework for BVS in large datasets using a bi-directional D\&C approach (Algorithm \ref{alg:my_algorithm}), which extends existing single-directional splitting methodology (\citep{consensus, liang_2015}) into a two-phase procedure. Such an approach offers extra flexibility in that the amount of row and column splitting can be adjusted to adapt to various datasets. The bi-directional splitting framework has introduced three new parameters: $L$ and $K$ representing the number of row and column splits, respectively, and $r$ which controls how many variables move from the first phase to the second. Furthermore, this work has also proposed an original covariate splitting technique (Equation \ref{eqn:split_example}) called `covariate shuffling' which aims to address the loss of joint covariate information incurred by splitting. \\

Experiments in Section \ref{sec:simulations} provided a more in-depth understanding of the bi-directional splitting framework's performance and highlighted the benefits of such an approach. First, initial experiments showed that lower values of $L$ and higher values of $K$ (Section \ref{sec:LK_r_training}), resulting in batches with more rows and fewer columns, are preferred. It was also shown in Section \ref{sec:training_comparison} that the bi-directional splitting framework yielded better predictive performance and, in a majority of cases, quicker computation time compared to SAM and CMC. Shuffling the covariates produced longer computation times on average than the simple splitting approach, but it also proved to be the more desirable option for moderate to highly correlated datasets, as shown in Section \ref{sec:corr_sim} . \\

Application of the proposed framework to the H3N2 dataset demonstrated its ability to perform variable selection in a challenging high-dimensional setting. The bi-directional splitting algorithm identified 23 candidate antigenic sites, of which the majority were located within known antigenic regions of the haemagglutinin protein or had previously been implicated in antigenic evolution \citep{h3n2}. In particular, several selected sites corresponded to substitutions identified through experimental studies, providing biological support for the variables selected by the method \citep{koel2013substitutions}. Furthermore, a small number of selected sites did not coincide with previously reported antigenic regions, suggesting potential avenues for future biological investigation. These results demonstrate that the proposed framework can scale to real-world datasets that are computationally challenging for traditional Bayesian variable selection approaches while still producing scientifically meaningful selections. \\

Several avenues for future work arise from the methodology developed in this paper. In particular, further investigation of alternative strategies for partitioning the covariates may lead to improvements in variable selection performance. More broadly, there is scope to develop more integrated and efficient approaches to combining information across batches. The flexibility of the proposed framework makes it well suited to such developments and to applications across a variety of fields and data structures.

\newpage

\section*{Acknowledgements}
The authors would like to thank William T. Harvey for his help with the H3N2 amino acid site identifications.

\bibliographystyle{plainnat}
\bibliography{bvs_split.bib}

\end{document}